\documentclass[10pt]{article}
\usepackage{amsfonts}
\usepackage{amsmath}

\newcommand{\be}{\begin{equation}}
\newcommand{\ee}{\end{equation}}
\newcommand{\bea}{\begin{eqnarray}}
\newcommand{\eea}{\end{eqnarray}}

\newtheorem{theorem}{Theorem}

\begin{document}

\title{Control aspects of holonomic quantum computation}
\author{Dennis Lucarelli  \\ 
The Johns Hopkins University \\ Applied Physics Laboratory \\ 
Laurel, Maryland 20723-6099} 
\date{}
\maketitle
\begin{abstract}
A unifying framework for the control of quantum systems with non-Abelian holonomy is presented.  
It is shown that, from a control theoretic point of view,  holonomic quantum computation can be 
treated as a control system evolving on a principal fiber bundle.  An extension of methods
developed for these classical systems may be applied to quantum holonomic systems to obtain insight 
into the control properties of such systems and to construct control algorithms for two established 
examples of the computing paradigm. 
\end{abstract}

\section{Introduction}Geometric phases have long been a source of fascination and insight
into classical and quantum physical theories \cite{phase}.  In recent years, they have
proven to be useful in describing the dynamics and control of certain nonholonomic mechanical systems
with symmetry \cite{blochBOOK}.  Inspired by the appearance of geometric phases in biology,
engineers have sought to create motion in machines via cyclic variations in {\it shape space}.
These endeavors \nocite{krishnaprasad1, kelly1, radford,ostrBurd} 
\cite{krishnaprasad1}-\cite{ostrBurd} and the characterization of optimal trajectories
\nocite{ostrowski, mont_cat,mont_tour,koon} \cite{ostrowski}-\cite{koon} remain active areas of research.

Most recently, the quantum geometric phase has been realized as a way of
constructing logic gates in a quantum computer \cite{zan}.
Holonomic Quantum Computation (HQC) employs non-Abelian geometric
phases (holonomies) for the purpose of quantum information processing. 
Here we present a unified framework for the control of quantum holonomic 
systems with applications to quantum computing by casting the model
as a control system evolving on a principal bundle.  The integration of
control theoretic ideas into the HQC paradigm sharpens existing results
for these systems and reveals computational techniques for solving two
separate but related fundamental problems in quantum computing.
First, the well known conditions for determining universal quantum computation
must be translated to the holonomic framework.  The determination of universality, 
however, is existential in nature and  generally not constructive.  Quantum logic gate synthesis 
or constructive controllability is the process of determining from the system dynamics the 
construction or concatenation from available transformations some desired 
dynamic transformation of the state. This latter task is required for executing 
quantum algorithms.  Both of these issues have subtleties in the 
holonomic framework not encountered in the usual {\em dynamical} approach 
to quantum computing. 

Since the introduction of this novel approach to quantum computing \cite{zan},
there has been considerable interest from the research community in 
proposing physical systems capable of performing HQC
\nocite{pach1, jones, duan, recati, pach3, solinas, faoro, choi} \cite{pach1}-\cite{choi}
and exploring its mathematical foundations 
\nocite{pach4, pach2, fujii1, fujii2,fujii2.5, fujii3, fujii4, luc, nisk1, nisk2,  tanimura} \cite{pach4}-\cite{tanimura}.  
This paper draws on much of this work to 
provide new a characterization of the problem, introduce novel computational
techniques and present new results for two classes of control architectures
related to HQC.  In particular, we state simple conditions for determining
the holonomy group of a principal fiber bundle with connection.  These conditions,
well known and reported elsewhere \cite{mont_cat,radford,luc}, 
circumvent the difficulties in directly applying
the Ambrose-Singer theorem which was originally stated as the technique for determining 
universality of HQC.  For quantum holonomy groups of dimension greater than 3,
a direct application of Ambrose-Singer essentially neglects
the contribution of nested Lie brackets of horizontally lifted vector fields, however these
vector fields can provide new transformations available for manipulating quantum
information.  From the product bundle representing single qubit rotations and two
qubit interactions, conditions for universality can then be derived.  Having established 
the decisive condition for determining universality, we explore the difficult inverse
problem associated to constructing holonomies.  Namely, given a desired holonomy
what is the loop in parameter space that generates it ?  In principal, this information
is sufficient for the experimentalist to construct particular holonomies in the laboratory.  
For the the ${\bf CP}^n$ model, holonomic logic gate synthesis has been recently 
addressed with a numerical optimization scheme in Refs. \cite{nisk1,nisk2}.  
Moreover, Ref. \cite{nisk2} refines the method to minimize the length of the
loop in parameter space.  Since the parameters must be driven sufficiently
slowly for the adiabatic approximation to hold, minimizing length also minimizes
the time to construct the logic gate.  This criterion is perhaps relevant for combatting
decoherence.  Logic gate synthesis has also been treated analytically in 
Ref \cite{tanimura}, however these loops are characterized in the Grassmann manifold and not
in parameter space.  Characterizations of length minimizing loops can also be found in 
Refs. \cite{mont_cat, mont_tour}.

We apply the theory to two well studied models of HQC.  We provide a complete analysis of 
the so-called \emph{Optical Holonomic Computer} \cite{pach1,pach2,fujii1,fujii2,fujii2.5, fujii3,fujii4}.
We extend the results of Ref. \cite{luc} by carrying out the universality analysis and explicitly
characterizing a parametric loop that can be employed to construct an arbitrary two-qubit
logic gate.  These results surmount a negative result reported for this
model \cite{fujii1}.  In a similar manner, we use the Cartan decomposition of the unitary group to 
solve the constructive controllability problem for holonomic systems involving a conditional
Berry phase.  To the best of our knowledge, aside from the ${\bf CP}^n$ model, 
the two control models treated here encompass all proposed holonomic 
computing schemes. 

This paper is organized as follows: In Section 2, following \cite{fujii1} the geometry
of holonomic quantum computation is reviewed. Also in Section 2, we introduce 
the product bundle describing single qubit holonomies and two qubit 
interaction holonomies and state conditions for universal holonomic quantum computation. 
In Section 3, we introduce methods for solving the path ordered integral associated with
logic gate synthesis in the holonomic framework.  The main contributions of the paper 
are contained in Section 4, where we apply the theory of the previous sections. 

\section{Holonomic Quantum Computation}
If a quantum state undergoes adiabatic evolution subject to a
periodic Hamiltonian, it acquires a phase after one complete
cycle.  Berry's surprising discovery \cite{berry} was that, in
addition to the well known dynamical phase associated to the
evolution, there is a phase of purely geometric origin.  Berry's
phase was then understood as the holonomy or geometric phase
corresponding to a principal bundle with connection over a
parameter space \cite{simon}.  This phenomenon has been
generalized in a variety of ways, most notably to non-adiabatic
evolutions \cite{aa} and to degenerate systems possessing a
non-Abelian phase factor \cite{zee}.

\subsection{Preliminaries}
We construct a family of degenerate Hamiltonians parameterized by elements of a parameter space
$M$ that govern the quantum dynamics. To allow for the possibility of a countably infinite dimensional 
Hilbert space, we consider universal classifying bundles. For further details see \cite{fujii1, fujii2.5,nak, bohm} .
 
Let ${\cal H}$ be a separable (possibly infinite dimensional) Hilbert  space,  
and define the manifolds
\begin{align}
St_k({\cal{H}}) &= \{ V = ( | v_1\rangle , \, \dots \, , | v_k \rangle) \in {\cal{H}} \times \, \dots \,
\times {\cal{H}} \,\, | \,\, V^{\dag}V = {\bf 1} \}   \\
Gr_k ({\cal{H}}) &= \{ X \in B({\cal H}) \,\, | \,\, X^2=X \, , \, X^{\dag} = X \, , \,
\mathrm{tr}X = n \} \, .
\end{align}
\noindent where $B({\cal H})$ denotes the set of bounded linear operators on ${\cal H}$.
These manifolds are known as the (universal) Stiefel and Grassmann manifolds respectively.
The space $Gr_k$ is also known as a  {\em classifying space} and can be defined as
the union of Grassmann manifolds \cite{nak}
\be \label{classify}
Gr_k ({\cal H}) \equiv \bigcup_{n = k}^{\infty} Gr_{k,n}({\cal H}) \, .
\ee  
Denote this $U(k)$-bundle by $P_k \, .$  Note that when ${\cal H} \cong {\mathbb C} ^n $ 
and the system has a $k$-dimensional degeneracy, the bundle of interest is the more familiar 
$U(k)$-bundle $St_{k,n}({\mathbb C} ^n) \to  Gr_{k,n}({\mathbb C} ^n) $  
which can be written in terms of coset spaces as
\begin{equation}
\frac{U(n)}{U(n-k)} \to \frac{U(n)}{U(n-k) \times U(k)} ~.
\end{equation}
We continue with the infinite dimensional case with the understanding that the development 
specializes to this case when ${\cal H}$ is finite dimensional. 

Let $M$ be a finite dimensional parameter space and suppose the classifying map 
$\Pi_k: M \to Gr_k$  (to be defined below) is given.    Then form the pullback bundle $Q_k = \Pi_k^*P_k \, , $
\begin{equation}
\label{pull}
 \begin{array}{ccc}
             Q_k &
        \longrightarrow &
         St_k({\cal H})  \\
        \Big\downarrow  & & 
          \Big\downarrow    \\
M & \stackrel {\Pi_k}{\longrightarrow}& Gr_k ({\cal H}) \, . 
\end{array}
\end{equation}

Let $H_0$ be a Hamilonian with a $k$-dimensional degeneracy spanned by the orthogonal
basis $\{ | v_j \rangle_{j=1}^{k} \} $.  To simplify notation, let the
degenerate eigenvalue be 0.  In holonomic quantum 
computation, the degenerate subspaces of  $H_0$ encode the quantum information.
Suppose we have at our disposal a set of $U(k)$ unitary transformations 
\be
 \{ W_1(x), W_2(x), \dots , W_m(x) \}
 \ee 
parametrized by the base coordinate $x$.  These are the (exponentiated) analogues of 
{\em control Hamiltonians}.  
Setting 
\be 
{\cal U}_k(x) = \prod_j W_j(x)
\ee
we obtain the isospectral family of Hamiltonians given by
\be 
{\cal O}(H_0) \equiv {\cal U}_k(x) H_0 \, {\cal U}_k^\dag(x) \, .
\ee
In the adiabatic approximation, the adjoint orbit ${\cal O}(H_0)$ forms a family of Hamiltonians 
that govern the system since there are no energy level crossings.    The classifying map is then 
be defined as 
\be
\Pi_k(x)  \equiv  {\cal U}_k(x) \Big( \sum_{j = 1}^k |v_j\rangle  \langle v_j | \Big) {\cal U}_k^\dag(x)\, .
\ee

\subsection{Control systems on principal fiber bundles}

In general, let $Q$  be a principal fiber bundle with structure group $G$ over a base manifold $M \, .$ 
Recall that a {\em connection} on $Q$ defines a $G$-invariant distribution $ \mathbb{H}$ such that 
$T_q Q = {\mathbb H}_q \oplus {\mathbb V}_q \, , $ where 
${\mathbb V}_q \equiv T_q ( {\cal O}(q) ) \cong {\mathfrak g}$ (the Lie algebra of $G \, )\, .$
Alternatively, a connection can be characterized by an Ad-equivariant ${\mathfrak g}$-valued 
one-form $ {\cal A} $ on $Q$ such that ${\cal A} \cdot \xi_q = \xi \, ,$
where $\xi_q$ is the infinitesimal generator of the group action and $\xi \in {\mathfrak g} \, . $
The horizontal subspace at a point $q \in Q$  is then defined as the kernel 
${\mathbb H}_q = \{ v_q \, \, | \, \, {\cal A} \cdot v_q = 0 \, \}\, .$  The local connection one-form, 
$A$, is defined with respect to a {\em local section} $\sigma$ by $A \equiv \sigma^*{\cal A} \, .$ 
Using Ad-equivariance and the fact that ${\cal A}$ is the identity on vertical vectors, we can 
obtain the local connection form in terms of the base variables only \cite{nak, bloch}
\be \label{local_form}
{\cal A} \cdot {\dot q} = \mathrm{Ad}_g( g^{-1} \dot{g} + A(x) \cdot \dot{x} ) \, .
\ee
We note that the term $g^{-1} \dot{g}$  is in the Lie algebra $\mathfrak{g}$, by interpreting  $g^{-1} \dot{g}$
as the lifted action of  $g^{-1} $ on $\dot{g} \in T_g G \, .$  Restricting the connection 
$\cal{A}$ to act on horizontal vectors yields an equation for the evolution of the group elements given by
\be
 g^{-1} \dot{g}  = -  A(x) \cdot \dot{x}  \, .
\ee

Returning to the quantum setting, we note that the canonical connection on the bundle $Q_k$ is given by 
$A_{wz} \equiv {\cal U}_k^{-1} d \, {\cal U}_k\, ,$ The matrix elements of the connection form are given by 
\begin{equation} \label{WZ}
A_{x_\mu}^{\bar{v} v} \equiv \langle \bar{v} |
{\cal U}_k^\dag (x) \frac{\partial}{\partial x_\mu} {\cal U}_k
(x) | v \rangle.
\end{equation}
This is commonly known as the Wilczek-Zee connection \cite{zee}.

Assuming direct control over the base variables, we may interpret the quantum control system as a 
{\em control system evolving on a principal bundle} and write it locally as
\bea \label{CSPB}
g^{-1}\dot{g} &=& -A_{wz}(x)\cdot \dot{x} \\
   \dot{x} &=& u \nonumber
\eea
where $u$ is a vector of control inputs describing the controlled evolution in parameter space.

A formal solution to this system of equations corresponding to a particular path in parameter 
space is  given by the {\em path ordered} integral 
\be \label{product_integral}
{\bf P}\,  \exp \int_\gamma -A_{wz}\, dx \, . 
\ee

When $\gamma \,$  is a closed curve in $ M \, , $ then ${\bf P}\,  \exp \oint_\gamma -A_{wz}\, dx$ lies in $G$ and is known as the 
{\em holonomy} of $\gamma \, .$  It is well known that the set of all such group elements taken over the set of closed curves in $M$
is a subgroup of $G$ and is known as the {\em holonomy group}.  In {\em holonomic quantum computation},  quantum logic gates
are implemented by holonomies acting on the degenerate subspaces.  

\subsection{Universality} 
A control system evolving on a principal fiber bundle is said to be {\em locally controllable} if any group element 
can be implemented on the state of the system.  In the context of quantum computing,  a system with this property is 
said to be (exactly) {\em universal}.  This property is a fundamental requirement for building a 
quantum processor.  Loosely speaking, in the usual dynamical approach 
to quantum computing (as opposed to the geometric approach addressed here), the Lie algebra 
generated by the system Hamiltonian and the control Hamiltonians determines the universality of
the system.  For an $N$ qubit system, it is sufficient for the Lie algebra to span $\mathfrak{su}(2^N) \,. $
We now show how this condition translates to the holonomic framework.

Let $X^h$ denote the horizontal lift of a vector field $X$ on $M$.  This is the unique vector on $TQ$ such that 
$T\pi(X^h) = X$ where $\pi$ is the projection 
$\pi : Q \to M \, . $  Then the  {\em curvature} can be defined as a $\mathfrak{g}$-valued 2-form on $Q$ given by
\be
{\cal A} ( [ X^h_{i_2}, \,   X^h_{i_1} ]  ) = - {\cal F } (X^h_{i_2}, \,   X^h_{i_1}) \, 
\ee where $[ \, \cdot \, , \, \cdot \, ] $ denotes the Lie bracket on $TQ\, .$
Thus evaluating the curvature determines the vertical component of the Lie bracket of horizontally
lifted vector fields. 
Now let $f : Q \to \mathfrak{g}$  be an Ad-equivariant function on $Q$, then 
\bea
{\cal A} ( [ X^h, \, f ] ) &=& -d{\cal A}(X^h, f) + X^h({\cal A}(f )) + f ({\cal A}(X^h) )\\
&=& X^h f 
\eea since the function $f $ is $\mathfrak{g}$-valued and $d{\cal A}(\cdot, \cdot)$  is zero if either 
argument is vertical \cite{nak}.  Using the correspondence between covariant derivatives 
of the associated adjoint bundle and Lie derivatives of Ad-equivariant functions \cite{kn}, 
we obtain
\be\label{curv}
{\cal A} ( [ X^h, \, f ] ) = X^h f  =  D_{X^h} f \, . 
\ee
Now, the curvature itself  is an Ad-equivariant function on $Q$ \cite{kn}, so setting ${\cal F} = f $ and using  the previous
expression (\ref{curv}) to evaluate iterated Lie brackets of horizontally lifted vector fields, we can obtain the corollary to 
the well known {\em Chow-Rashevski} theorem from control theory.
\begin{theorem}(Ambrose-Singer-Chow-Rashevski)
The system (\ref{CSPB}) is locally controllable at $q \in Q$ if the curvature $F (X_{i_1} , X_{i_2})$ and all of its 
covariant derivatives $D_{X_{i_k}}  \cdots D_{X_{i_3}}  F (X_{i_1} , X_{i_2})$ evaluated at the point $x = \pi(q) $ span
the entire Lie algebra of $G$ . 
\end{theorem}
Following \cite{mont_tour}, we refer to the theorem as
{\em Ambrose-Singer-Chow-Rashevski} since it can be considered to be a corollary 
to the {\em Ambrose-Singer} theorem from the theory of holonomy \cite{kn}.   We note also
that we have stated the theorem in terms of base vector fields and the local curvature.    
All the necessary ingredients of the theorem, although not explicitly stated, can be found in \cite{kn}.
In fact, the {\em infinitesimal holonomy algebra} is spanned by elements of the form 
\be
X_{i_k}^h \cdot X_{i_{k-1}}^h  \cdots X_{i_3}^h \cdot {\cal F} (X_{i_1}^h , X_{i_2}^h) \, . 
\ee
We can then use the correspondence (\ref{curv}) to relate this to covariant derivatives of the associated adjoint bundle.  This
statement is used in our applications, since in some holonomic quantum computation problems the relevant holonomy
algebra does not span the entire Lie algebra.  However, it does contain {\em non-local} operations which together with
holonomies corresponding to {\em local} operations do indeed span the entire Lie algebra.  This is the usual 
local/non-local analysis often encountered in quantum information science.  

For the purposes of building a quantum processor, the quantum information is stored in the $\mathbb{C}^{2}$ 
vector bundle associated to $Q^2$ and single qubit rotations are performed by $SU(2)$ holonomies acting 
on the fiber $\mathbb{C}^{2} \, .$  Interactions among qubits are modeled as $SU(4)$ holonomies acting on the 
fibers of the vector bundle associated to $Q^4\, .$

Thus we may treat the control problems separately and form the product bundle (and its pullbacks)
\begin{equation}
\label{unibund}
 \begin{array}{ccccccc}
    \cdots & St_{2}({\cal H}_i) &
      \times  &
     St_{4}( \cdots \otimes {\cal H}_i \otimes {\cal H}_j \otimes \cdots ) &
      \times  &
     St_{2}({\cal H}_j ) & \cdots  \\
     &   \Big\downarrow  & &
      \Big\downarrow     & &
      \Big\downarrow  &  \\
  \cdots &   Gr_{2}( {\cal H}_i) &
       \times &
      (Gr_{4})^{\rm int}(\cdots \otimes   {\cal H}_i \otimes {\cal H}_j \otimes \cdots) &
       \times &
      Gr_{2} ({\cal H}_j ) & \cdots \quad . 
\end{array}
\end{equation}

To set notation, let 
\be
I_x =  \frac{1}{2}
\left( 
\begin{array}{lr}
0& \,\,1 \\ 1& \,\,0
\end{array} \
\right)  \quad 
I_y  =  \frac{1}{2}
\left(
\begin{array}{lr}
0&-i \\ i&0
\end{array} 
\right)  \quad 
I_z =  \frac{1}{2}
\left(
\begin{array}{lr}
1&0 \\ 0&-1
\end{array}
\right)  
\quad 
{\bf 1} =  
\left(
\begin{array}{lr}
1&0 \\ 1& 0
\end{array}
\right)  \, \, . 
\ee
We define the {\it local algebra} generated by the elements
\be \label{loc_alg}
I_{k1}= I_k \otimes {\bf 1} \quad I_{k2} = {\bf 1} \otimes I_k 
\ee
where $k \in \{ x\, , y\, , z \} \, $ 
The local algebra is the Lie algebra corresponding to the {\em local group} $SU(2) \otimes SU(2) \, .$

To conclude  exact universality (controllability) of the system, one should  compute the {\em control Lie algebra}  with the 
constituent holonomy algebras.  For example in the two-qubit system, compute the Lie algebra generated
by $\mathfrak{su}(2)\otimes {\bf 1} \, , \, {\bf 1}\otimes\mathfrak{su} (2)$ and the interaction holonomy algebra $\mathfrak{hol}_{int}$
 associated with the bundle   $\mathrm{St}_{4}(  {\cal H}_i \otimes {\cal H}_j )  \longrightarrow  
(Gr_{4})^{\rm int}( {\cal H}_i \otimes {\cal H}_j  ) \, . $  In the generic case, the control Lie algebra
will generate $\mathfrak{su}(4)$ provided that $\mathrm{Hol}_{int}$ is not isomorphic to the local group or is
trivial \cite{lloyd1,nik}.

\begin{theorem}
The two-qubit holonomic system is  exactly universal if the Lie algebra generated by the local algebra and $\mathfrak{hol}_{int}$
spans $\mathfrak{su}(4) \, .$
\end{theorem}

 \section{Constuctive Controllability}
Having established conditions for determining universality in HQC, we now present various ways 
of solving or approximating the solution to the path ordered integral
arising for the differential equation defining the group displacement 
\be \label{group_de}
g^{-1}\dot{g} = - A(x) \cdot \dot{x} \, .
\ee
For control systems on principal bundles, this equation describes the group transformation obtained
from a controlled cyclic variation of the parameters in the base manifold.  Recall, that we assume direct access
and complete controllability over the base variables.  We endeavor to ascertain the desired group transformation
resulting from a particular choice of loop in the base space.  This is the notion of constructive controllability in the 
context of a control system on a principal bundle.  In HQC these procedures provide explicit methods for logic
gate synthesis. This requires dealing with the path ordered integral 
obtained from (\ref{group_de}).

\subsection{Path Ordered Integral}
We define the path ordered integral as a product integral.  
Let $\gamma$ be a curve in the base manifold $M\,$ and let $x^\mu$ be local coordinates.  We may express
the local connection form $A$ in terms of coordinates as
\be
A(x) = A_\mu (x) dx^\mu \, . 
\ee
 
The curve $\gamma$ is parameterized by an intrinsic parameter $s$, which is naturally considered to be 
time.   In terms of $s$, $A$ takes the form
\be
A_\mu(x(s))d x^\mu = A(s)ds 
\ee
where
\be
A(s) \equiv A_\mu (x(s))\frac{d x^\mu (s)}{ds} \, .  
\ee

Let $[s_0, s_T]$ be a real interval over which the  curve $\gamma$ is defined.   Consider a partition of the interval
$P = \{s_0, s_1, \dots, s_n \} $ such that $\Delta s_k = s_k  - s_{k-1} $ and $s_n = s_T \, . $  
Then path ordering operator may be defined as 
\be \label{product_integral}
{\bf P}\,  \exp \int_\gamma -A (s) \, ds   \equiv  \lim_{n\to \infty} \prod_{s_0}^{s_n} \exp \left(-A(s_k)\Delta s_k \right) \, .
\ee
This definition clearly shows the dependence on the ordering of the exponentials and the difficulties associated
with its solution, given that we are naturally interested in the case where the relevant group is non-Abelian. 
When $G$ is Abelian, then one can directly integrate  the connection coefficients and apply the usual
exponential operator.   

\subsection{Abelian Substructures}
A common technique in holonomic quantum computation for tackling the integral (\ref{product_integral}) is 
to restrict the class of loops and exploit Abelian substructures in the connection components  \cite{pach1,pach2}.    
The strategy is briefly described as follows.   Choose a particular 2-manifold of $M$ spanned by the coordinates $(\sigma, \tau)$ such
that the associated connection  components commute, that is $[A_\sigma, A_\tau] = 0 \, $ but for which the local curvature
form is not identically zero.   For these restricted loops the path ordering in (\ref{product_integral}) can be avoided and the 
line integral 
\be 
\oint_\gamma A(x) dx  = \oint_\gamma  \left( A_\sigma d \sigma   + A_\tau  d \tau \right)
\ee
can be integrated directly and exponentiated. 

Alternatively, one can use a non-Abelian Stokes theorem \cite{karp} for evaluating holonomies corresponding to curves lying in 
a 2-submanifold of parameter space.

\subsection{Averaging}
The exact results of the previous section were accompanied by restrictions on the set of loops available to 
the controller or by exploiting Abelian substructures in the connection components.  Here we review 
local approximations that can be used for any system evolving on a principal bundle. 

Approximate control algorithms have been developed for left invariant control systems on Lie groups of the form
\be \label{LINV}
g^{-1} \dot{g} = \epsilon U(t)
\ee
where $\epsilon$ is a (small) parameter and  $U(t) = T_\alpha u^{\alpha}(t) $ for a basis $\{ T_\alpha \}$ of $\mathfrak{g} \, $
 \cite{krishnaprasad2}.

A Magnus expansion is employed for a representation of the solution
\be 
g(t) = g(0)\exp(\xi(t))
\ee  given by
\bea \label{average}
\xi(t) = \epsilon \int_0^t U(\tau)d\tau &+& \frac{\epsilon^2}{2}\int_0^t [ \tilde{U}(\tau), U(\tau) ] d\tau   \nonumber \\  
 &+& \frac{\epsilon^3}{4} \int_0^t \left[ \int_0^\tau [ \tilde{U}(\sigma), U(\sigma) ] d\sigma , U(\tau)  \right]  d\tau  + \dots 
\eea
where  $\tilde{U}(t)$ is the effective input ``averaged" over the time period \cite{krishnaprasad2} .  

This expansion has been generalized for systems evolving on principal fiber bundles $\cite{radford} \,  .$  
Let $\gamma : [ 0,T] \to M$ be a  closed curve in the base space parameterized by $x \in M $ , then it is shown in  \cite{radford} 
that the holonomy associated to $\gamma$ can be locally approximated by 
\be 
g(T) = g(0)\exp(\xi(\gamma))
\ee  where
\be\label{average}
\xi(\gamma) = -\frac{1}{2} F(X_i, X_j) \int_\gamma dx_i dx_j + \frac{1}{3} D_{X_i} (F(X_j, X_k)) \int_\gamma dx_i dx_j dx_k + \dots \quad .
\ee
Here $F(X_i, X_j)$ is the local curvature form evaluated on the base coordinate vectors $X_i = \frac{\partial}{\partial_{x_i}} \, $
evaluated at $\gamma(0) \,  , \, \, D_{\frac{\partial}{\partial_{x_i}} }$ is the covariant derivative of the curvature along the base coordinate vector
$\frac{\partial}{\partial_{x_i}} $ and the area integrals are defined by
\be
{\cal I}_{x_ix_jx_k} = \int_\gamma dx_i dx_j dx_k \equiv \int_0^T \int_0^{t_k} \int_0^{t_j} \dot{x}^i(t_i) dt_i \dot{x}^j(t_j) dt_j \dot{x}^k(t_k) dt_k  \, .
\ee
Higher order terms are given by higher order covariant derivatives of the curvature.  This is plausible given 
expression (\ref{average}) and the fact that iterated Lie brackets of horizontally lifted vector fields appear as covariant
derivatives of the curvature.

\section{Applications}
In this section we apply the results of the preceding sections.  We first review the 
 ${\bf CP}^n$ model of quantum holonomy.  This was the original system discovered by Wilczek and Zee \cite{zee} and
 subsequently proposed as a model for HQC. We then consider two very different models of  quantum 
 holonomic systems.  Holonomic quantum computation with  squeezed coherent states has a rich interaction holonomy group 
 that can be exploited to obtain contructive controllability algorithms.  On the other hand, quantum computation based on the 
 conditional phase shift has become the dominant control  strategy for a wide range of holonomic quantum computing schemes.  

\subsection{The ${\bf CP}^n$ Model}
The ${\bf CP}^n$  \cite{zan} model gives a concrete example illustrating how non-Abelian holonomies can occur
in highly degenerate systems.  In this model, we assume that the Hilbert space ${\cal H}$ is finite dimensional
from the outset.  That is, we have the isomorphism ${\cal H} \cong \mathbb{C}^{n+1} = \{ | \alpha \rangle \}^{n+1}_{\alpha =1} \, .$  
We further assume an $n-$dimensional degenerate subspace with eigenvalue 0.  We may write the degenerate Hamiltonian $H_0$ as
\be
H_0 = \epsilon | n + 1 \rangle \langle n+1 |  =  
\left(
\begin{matrix}
0&0&\dots &0 \\
0 & \ddots &0& 0\\
\vdots & & \ddots& \\
0&0&0& \epsilon
\end{matrix}
\right) \, . 
\ee

Let $\{ {\cal T}_{j,n+1}(x)\}_{j=1}^{l}$  denote a basis of $\mathfrak{u}(n) $ parameterized by $x \in M $ embedded in $\mathfrak{u}(n+1)$ and let 
${\cal U}_n = \prod_j^l \exp({\cal T}_{j,n+1}(x)) \, . $ Given these control operations, it is perhaps not surprising that the holonomy group 
can be shown to be $U(n) $ by considering the curvature coefficients only (and not its covariant derivatives) \cite{zee,zan}.  We have the isomorphism 

\be\label{class_iso}
{\cal O} (H_0) \cong \frac{U(n+1)}{U(n) \times U(1)} \cong \frac{SU(n+1)}{U(n)} \cong {\bf CP}^n \, . 
\ee

This system requires control over $2 n = \mathrm{dim}_{\mathbb R}  {\bf CP}^n $  
real parameters to control an $n$-level  system. For high dimensional systems, this may be an unrealistic requirement.  

\subsection{Squeezed Coherent States}
In this section we revisit the mathematical foundations of {\em holonomic quantum 
computation with squeezed coherent states} .  There is considerable literature 
already on this model \cite{pach1,pach2,fujii1,fujii2,fujii3,fujii4}, 
here we exploit the  methods of geometric control.  
Originally, this model was proposed in the context of quantum optics \cite{pach1} with displacers and squeezers
operating on coherent laser beams in a non-linear Kerr medium and thus known as the Optical Holonomic Computer. 
However, other physical systems have quantum states that may be displaced and squeezed.  As far as the control analysis 
is concerned, these systems are identical.  Pachos has recently adapted the model to perform trapped ion quantum computation \cite{pach3}. 
We, therefore, refer to this model generically as  {\em holonomic quantum computation with squeezed coherent states} .  

\subsubsection{Harmonic Oscillator}
Recall that  the commutation relations of the creation, annihilation and number operator $N \equiv a^\dag a \, ,$ are given by
\begin{equation}
[N,a^\dag] = a^\dag \, , \quad [N,a]= a \, , \quad [a,a^\dag] = {\bf 1}  \, .
\end{equation}
The underlying Hilbert space ${\cal H}$  is a Fock space and takes the  form 
\be
{\cal H}= \{\, | n \rangle \, , \, \alpha \in {\bf N} \cup {0} \, \} .  
\ee
The creation and annihilation operators act on ${\cal H}$ according to 
\begin{equation}
 a | n \rangle = \sqrt{n} | n-1 \rangle \, , \, \,  a^\dag | n \rangle = \sqrt{n+1} | n+1 \rangle 
\, , \,\, a | 0 \rangle = 0 .
\end{equation}
Thus $a$ and $a^{\dag}$ create and destroy quanta.  
 
Since we are interested in the two-qubit system, we will use the subscript $i$ to 
distinguish the creation and annihilation corresponding to the $i-th$ field of the harmonic 
oscillator.  That is, we set $N_i = a_i^{\dag} a_i $ and 
\begin{eqnarray}
&a_1 = a \otimes {\bf 1}\, , \quad  a_1^\dag = a^\dag \otimes {\bf 1} \, , \\
&a_2 = {\bf 1} \otimes a \, , \quad   a_2^\dag = {\bf 1} \otimes a^\dag \, .
\end{eqnarray}

To provide a concrete example, we will use the degenerate Hamiltonian
\begin{equation}\label{H1}
H^i = N_i (N_i -1)
\end{equation}
to encode the $i$-th
qubit in the degenerate subspace $\{ | 0_i \rangle \, , \, |  1_i \rangle \} \, $
and 
\begin{equation}\label{H12}
H^{12} = N_1(N_1-1) + N_2(N_2-1)
\end{equation}
to obtain controlled interactions on the computation basis 
$\{\, | 00 \rangle,  | 0 1 \rangle ,  | 10 \rangle,  | 1 1 \rangle \, \} 
$where $| i j \rangle = | i \rangle \otimes | j \rangle \, .$   In the optics context, 
this Hamiltonian corresponds (up to a constant)  to placing lasers in a non-linear Kerr medium 
\cite{pach1,pach2}.   However,  the form of the degenerate Hamiltonian does not 
affect the control analysis.  For our purposes, it is used only to encode the quantum information.
With slight modification of some constants, the results in this section apply to the trapped ion
model proposed in \cite{pach3}.

\subsubsection{Single Qubit}
We consider first single qubit rotations.   Consider the eigenvalue problem for a single creation
operator. The state $| \alpha \rangle $ can be written in terms of the basis $\{| n  \rangle \} \, , $
\be 
| \alpha \rangle = e^{- |{\alpha}|^2/2 }\sum_{n=0}^{\infty} \frac{\alpha^n}{\sqrt{n !}}| n \rangle  \,  
=  e^{- |{\alpha}|^2/2 }\sum_{n=0}^{\infty} \frac{\alpha^n a^{\dag n }}{n !} | 0 \rangle  . 
\ee
Which is equal to 
\be
| \alpha \rangle = e^{- |{\alpha}|^2/2 } e^{\alpha a^{\dag}} | 0 \rangle
\ee
and allows for the definition of the {\em displacement operator} 
\be
| \alpha \rangle = D(\alpha) = e^{- |{\alpha}|^2/2 } e^{\alpha a^{\dag}} e^{-\alpha^*  a} | 0 \rangle \, . 
\ee
Note that the introduction of $e^{-\alpha^* \,   a}$ does nothing since $a | 0 \rangle = | 0 \rangle \, . $  By using the
Campbell-Baker-Hausdorff formula and noting that $[ \alpha a^{\dag} , -\alpha^* a ] = |{\alpha}|^2 \, , $ we may
write
\be
| \alpha \rangle = D(\alpha) | 0 \rangle = e^{\alpha a^{\dag} - \alpha^* a} | 0 \rangle \, . 
\ee
We see that the displacement operator creates a coherent state $| \alpha \rangle$ from the vacuum state $ | 0 \rangle \, . $
In HQC with squeezed coherent states, the displacement operator will be a control operator.  The  other transformation
we have at our disposal is the {\em squeezing operator}, $S(\beta) \, , $  defined by
\be
S(\beta) = e^{\beta \Lambda_+ - \beta^* \Lambda_-}
\ee
where
\be
\Lambda_+ \equiv  \frac{1}{2} a^{\dag 2} \quad  \Lambda_- \equiv  \frac{1}{2} a^{2} \, .
\ee
If we define,
\be
\Lambda_3 \equiv  \frac{1}{4} (  a a^{\dag} + a^{\dag} a )
\ee
then we have the commutation relations,
\be
[ \Lambda_3, \Lambda_+ ] = \Lambda_+ \quad [ \Lambda_3, \Lambda_- ] = -\Lambda_-  \quad [ \Lambda_+, \Lambda_- ] = 2 \Lambda_3 \, . 
\ee 
These are the commutation relations for  $\mathfrak{su}(1,1)$;  thus we see that the squeeze 
operator is a representation of the non-compact group $SU(1,1) \, . $  With these two unitary transformations, we form the product
\be
{\cal U}_2(\alpha, \beta) = D(\alpha)S(\beta)
\ee
and the isospectral family of Hamiltonians
\be
{\cal U}_2(\alpha, \beta) H^i {\cal U}_2^{\dag}(\alpha, \beta) \, . 
\ee
The holonomy group for the single qubit system has been shown to be
$U(2)$ \cite{pach1,pach2,fujii1,fujii2}.  
Thus we have complete control over the single qubit.  
  
\subsubsection{Two-qubit}
To obtain universality over the entire quantum register it suffices to show non-trivial $U(4)$ transformations on the 
computational basis and check the control Lie algebra.  
Analogously to the single qubit case, we employ displacement and squeeze operators as our control operations.  Let 
\be
J_+ = {a^{\dag}}_1a_2\, , \quad J_- = {a^{\dag}}_2a_1 \, , \quad J_3 = \frac{1}{2}({a^{\dag}}_1a_1 - {a^{\dag}}_2a_2) \, . 
\ee
These generate $SU(2) \, $ with the commutation relations, 
\be
[J_3, J_+] = J_+ \, , \quad [J_3, J_-] = -J_-  \, , \quad [J_+, J_-] = 2 J_3 \, \, . 
\ee 
The two-mode displacement operator is defined as 
\begin{equation}
N(\xi) = \mathrm{exp} \, (\xi a_1^{\dag}a_2 - \bar{\xi}a_1a_2^{\dag}) \, . 
\end{equation}
Similarly, we may define the two-mode squeeze operator as a representation of $SU(1,1) \, . $  Let
\be
K_+ = {a^{\dag}}_1{a^{\dag}}_2 \, , \quad K_- = a_1a_2  \, , \quad K_3 = \frac{1}{2} ( {a^{\dag}}_1a_1 + {a^{\dag}}_2 a_2 ) , 
\ee
and
\be 
[K_3, K_+] = K_+ \, , \quad [K_3, K_-] = -K_-  \, , \quad [K_+, K_-] = -2 K_3 \, \, . 
\ee 
The two-mode squeeze operator is defined as
\be
M(\zeta) = \mathrm{exp} \, (\zeta a_1^{\dag}a_2^{\dag} - \bar{\zeta} a_1a_2) \
\ee
where $\xi \, , \, \zeta \, \in {\bf C} . $  Set 
\be
{\cal U}_4 = N(\xi)M(\zeta). 
\ee
Setting  $\zeta = r_2 e^{ i \theta_2} \, \mathrm{and} \, \xi= r_3 e^{i\theta_3} \,$ and we obtain the two-qubit
connection coefficients \cite{pach2} listed in  Appendix A.  The interaction holonomy algebra spans 
$\mathfrak{su}(2) \times \mathfrak{su}(2) \times \mathfrak{u}(1) $  \cite{luc} (also listed in Appendix B). 
Higher order covariant derivatives do not yield independent group directions.  The matrices in $\mathfrak{hol}_{int}$ sit in 
$\mathfrak{u}(4)$ in a manner that allows for non-local $U(4)$ transformations on the computational basis.   By the
reduction theorem for connections \cite{kn}, the connection is reducible to a $su(2) \times
su(2) \times u(1)$-valued connection and we may reduce the total space to
$\mathrm{St}_{2,4}({\cal H}_1 \otimes {\cal H}_2) ~. $  To determine the reduced base manifold, we
form the quotient
\begin{equation}
\frac{U(4)}{SU(2) \times SU(2) \times U(1)}  \cong  \frac{SU(4)}{SU(2) \times SU(2)}
\cong {SGr}_{2,4} \equiv ({Gr}_{2,4})^{\rm int} ~.
\end{equation}
In a similar manner, we can reduce the bundles ${St}_2 ({\cal
H}_i) \mapsto {Gr}_2 ({\cal H}_i)$ corresponding to the single
qubit rotations. The $U(2)$ holonomies act in the product space
${\cal H}_1 \otimes {\cal H}_2 $ as $U(2) \otimes {\bf 1}$ and $
{\bf 1} \otimes U(2) .$  The bundles reduce to 
$ {St}_{2,4}({\cal H}_1 \otimes {\cal H}_2) \mapsto {Gr}_{2,4}({\cal H}_1 \otimes {\cal H}_2). $  
For the full two-qubit system, we have the reduced product bundle
\begin{equation}
\label{unibundreduced}
 \begin{array}{ccccc}
      {St}_{2,4}({\cal H}_1 \otimes {\cal H}_2 ) &
      \times  &
       {St}_{2,4}({\cal H}_1 \otimes {\cal H}_2) &
      \times  &
       {St}_{2,4}({\cal H}_1 \otimes {\cal H}_2 )  \\
        \Big\downarrow  & &
      \Big\downarrow     & &
      \Big\downarrow    \\
       {Gr}_{2,4}({\cal H}_1  \otimes{\cal H}_2) &
       \times &
      ( {Gr_{2,4}})^{\rm int}({\cal H}_1 \otimes {\cal H}_2) &
       \times &
      Gr_{2,4} ({\cal H}_1 \otimes{\cal H}_2 ) \, \, . 
 \end{array} 
\end{equation} 

\subsubsection{Control Algebra}

To be complete, we will now demonstrate that all of $SU(4)$ may be obtained from the single qubit rotations and the two-qubit
transformations above.  Of course, as we have mentioned earlier, this is generically true 
provided the two-qubit holonomy group  is not isomorphic to the local group.   
Nonetheless, it is useful to go though the computations.

From the single qubit analysis,  we know that we can perform local transformations of the form $SU(2)\otimes SU(2) \, .$  
To simplify matters further, we use linear combinations of the the two-qubit curvature forms and covariant derivatives 
and consider only 
\be
\{ F_{r_2 r_3} \, , F_{r_2 \theta_3} \, , F_{r_3 \theta_3} ,  D_{\frac{\partial}{\partial \theta_2}} F_{r_{2}\theta_{2}} \, , 
\tilde{ D}_{\frac{\partial}{\partial r_2}} F_{r_{2}\theta_{2}} \, \}
\ee
where 
\begin{equation}
\tilde{D}_{\frac{\partial}{\partial r_2}} F_{r_{2}\theta_{2}} \, =  \,
\left(
\begin{matrix}
0 & 0 & 0 &  -e^{-i\theta_2} \\ 0 & 0 & 0 & 0 \\
0 & 0 & 0 & 0 \\ -e^{i \theta_2} & 0 & 0 & 0
\end{matrix}
\right)
4i\sinh 2r_2 \quad . 
\end{equation}
From this set of matrices and the local algebra (\ref{loc_alg}), 
we may build a set of holonomic transformations spanning $\mathfrak{su}(4)\, . $
After taking iterated brackets from these sets, we find that one choice of spanning elements is given by
\be
\mathfrak{su}(4) = \{ {\cal C}_1 \cup {\cal C}_2 \cup {\cal C}_3\} 
\ee
where $\,  {\cal C}_1 =  \{ I_{x1} ,  I_{y1}, I_{z1}, I_{x2}, I_{y2}, I_{z2}, \}  \, $ 
\noindent and
\bea
&&{\cal C}_2\ = \{ F_{r_2 r_3} ,  F_{r_2 r \theta_3} ,  D_{\frac{\partial}{\partial \theta_2}} F_{r_{2}\theta_{2}} , \tilde{D}_{\frac{\partial}{\partial r_2}} F_{r_{2}\theta_{2}} \}  \nonumber \\
&&{\cal C}_3 \ = \left\{ \left[ \,  I_{x1} \, , \, F_{r_2 r_3} \, \right]  , \left[  I_{x1} \, , \, F_{r_2 \theta_3} \right] ,  \left[ I_{x2} \, , \, F_{r_2 r_3}  \right] , \left[  I_{x2} \, , \, F_{r_2 \theta_3} \right] , \Big[  [\,  I_{x1} \, , \, F_{r_2 r_3} \, ]    \, , \, I_{x2}\Big]   \right\} \, .  \nonumber
\eea
Please see Appendix C for the matrix representation of these elements.  

\subsubsection{An approximate holonomy in the Cartan subalgebra of $\mathfrak{su}(4)$}
In the preceding section, we showed that it is indeed  {\em possible} to create holonomic transformations spanning the full unitary group on
two qubits.  This was not a constructive procedure.   In this section, we show that by using a combination of the methods in the previous
sections, we can solve the logic gate synthesis problem completely.  We use the local expansion of the holonomy procedure to 
construct an element in the Cartan subalgebra of $\mathfrak{su}(4)$ and use the Cartan decomposition of $SU(4)$ to obtain the result.

The Cartan decomposition of the unitary groups is a useful technique that has been used for constructing quantum control algorithms 
\cite{dalles3,khan2}, deriving time optimal control laws for quantum spin systems \cite{khan1} and 
understanding the entanglement content of 2-qubit unitaries \cite{cirac,linden}.  Here we review the 
decomposition for the purposes of constructing control algorithms.  

Let $K$ denote a closed and compact subgroup of a Lie group $G$.  Assume that $\mathfrak{g}$ admits a vector space decomposition 
\begin{equation}
\frak{g} = \frak{k} \oplus \frak{p}
\end{equation}
where $\frak{k}$ is the Lie algebra of $K$ and $\frak{p}$ is vector space orthogonal to $\frak{k}$ with respect to a 
bi-invariant metric $ < \cdot , \cdot >$ on $\frak{g}$. Further assume that $ \frak{k} \, \, \mathrm{and} \, \, \frak{p}$ satisfy the following commutation relations
\begin{equation}\label{cartan_comm}
[\mathfrak{k},  \mathfrak{k} ]  \subset   \mathfrak{k}  \quad [\mathfrak{p}, \mathfrak{p} ] \subset \mathfrak{k} \quad [\mathfrak{p} , \mathfrak{k} ] \subset \mathfrak{p}  \, .
\end{equation}
We refer to a this decomposition as a Cartan decompostion of the Lie algebra $\mathfrak{g}$. 

Let $\mathfrak{a}$ denote a maximal Abelian subalgebra contained in $\mathfrak{p}$. The algebra $\mathfrak{a}$ is often called the 
{\em Cartan subalgebra} of $\mathfrak{g} \, .$ Then one can write $G$ as
\be
G = KAK
\ee
where $A = \exp(\mathfrak{a}) \, .$  

In a two-qubit system,  interactions among the qubits are modeled by the  products
\be
I_{kl} = 2 I_k \otimes I_l 
\ee
where $k \, , l \in \{ x\, , y\, , z \} \, $ For example,  
\be
I_{yy} = \frac{1}{2}
\left(
\begin{matrix}
0&0&0&-1 \\
0&0&1&0\\
0&1&0&0\\
-1&0&0&0
\end{matrix} 
\right) \quad . 
\ee
The Cartan decomposition of $\mathfrak{su}(4)$ is given by
\bea
\mathfrak{k} &=& i\{ I_{x1}, I_{y1}, I_{z1}, I_{x2}, I_{y2}, I_{z2} \} \\
\mathfrak{p} &=& i\{  I_{xx}, I_{xy}, I_{xz}, I_{yx}, I_{yy}, I_{yz}, I_{zx}, I_{zy}, I_{zz} \} \\
\mathfrak{a} &=& i\{  I_{xx}, I_{yy}, I_{zz} \} \, .
\eea
Thus we can write any $g \in SU(4)$ as 
\be
  g = K_1 \exp({ -i\phi_1 I_{xx} -i\phi_2 I_{yy} -i \phi_3 I_{zz}})K_2
\ee    
 where $\phi_{j} $ is a real parameter and $K_j \in SU(2) \otimes SU(2) \, . $

By inspection of the two-qubit curvature forms and their covariant derivatives, it seems possible that $I_{yy}$ can be obtained
by linear combinations the elements, 
\be
\left\{ F_{r_2\theta_2} , F_{r_2 \theta_3} , D_{\frac{\partial}{\partial r_2}} F_{r_{2}\theta_{2}} , 
D_{\frac{\partial}{\partial r_2}} F_{r_{2}\theta_{3}} \right\} \, . 
\ee Equivalently, $I_{yy}$ is contained in the real span of 
\be
\left[ {\frac{\partial^{\, h}}{\partial x_k}} , \left[ {\frac{\partial^{\, h}}{\partial x_{k-1}}} \, , \, \dots \left[ {\frac{\partial^{\, h}}{\partial x_2}} ,  {\frac{\partial^{\, h}}{\partial x_{1}}} \right] \right] \cdots  \right] 
\ee where $x \in \{ r_2, \theta_2, \theta_3 \} \, . $

We therefore choose a candidate loop, $\gamma^* \, ,$ of the form
\bea
\theta_2(t) &=& \theta_2(0) + \Theta_2 \sin(t ) \\
r_2(t) &=& r_2(0) + R_2 \cos( t ) - R_2 \\
\theta_3(t) &=& \theta_3(0) + \Theta_3 \sin(n  t ) \, , \quad  n \neq 1 \\ 
r_3 &=& constant 
\eea
with the parameters $ \left\{n,  r_2(0), \theta_2(0), \theta_3(0), R_2, \Theta_2 ,\Theta_3  \right\} $ to be determined.
We compute the integrals appearing in the expansion (\ref{average}), with the period $T = 2 \pi \, $ and choose
some parameters to yield the expressions,   
\begin{eqnarray}
F_{r_2\theta_2}  \cdot {\cal I}_{r_2\theta_2}  &=& 
\left(
\begin{matrix} 0 & 0 & 0 & 0 \\ 0 & 1 & 0 &  0 \\
0 &  0 & 1 & 0 \\ 0 & 0 & 0 & 2 \\
\end{matrix}
\right)
2 i \,  \sinh 2r_2(0) \cdot  R_2 \Theta_2 \pi  \\
  \\ 
F_{r_2\theta_3}\Big{|}_{\theta_3(0) = \pi} \cdot {\cal I}_{r_2\theta_3}  &=& 
\left(
\begin{matrix}
0 & 0 & 0 & 0 \\
0 & 0 & -1&0\\ 0 & -1& 0 & 0 \\ 0& 0 & 0 & 0 \\
\end{matrix}
\right)  \,
i \sin 2r_3  \sinh 2r_2(0)  \cdot \frac{R_2 \Theta_3 \sin 2n\pi}{n^2-1} \\   
D_{\frac{\partial}{\partial r_2}} F_{\theta_{3} r_2}\Big{|}_{\theta_3(0) = \pi} \cdot {\cal I}_{r_2 \theta_3 r_2}   &=&    
\left(
\begin{matrix}
0 & 0 & 0 & 0 \\
0 & 0 & -1&0\\ 0 & -1& 0 & 0 \\ 0& 0 & 0 & 0 \\
\end{matrix}
\right) 2 i\sin 2 r_3  \cosh 2 r_2(0)  \cdot \frac{-6R^2_2  \Theta_3 \sin 2n\pi}{n^4-5n^2+4}   \\
D_{\frac{\partial}{\partial r_2}} F_{r_{2}\theta_{2}} \Big{|}_{\theta_2(0) = 0 } \cdot  {\cal I}_{r_2 r_2  \theta_2}  &=& 
\left(
\begin{matrix}
0 & 0 & 0 &  -1 \\ 0 & 0 & 0 & 0 \\
0 & 0 & 0 & 0 \\ -1 & 0 & 0 & 0
\end{matrix}
\right)
4i\sinh 2r_2(0)  \cdot 2 \Theta_2 R^2_2 \pi 
\,  \nonumber \\ && \nonumber \\  \,
&& \quad + \, \,  \left(
\begin{matrix} 0 & 0 & 0 & 0 \\ 0 & 1 & 0 &  0 \\
0 &  0 & 1 & 0 \\ 0 & 0 & 0 & 2 \\
\end{matrix}
\right)
4i \cosh 2r_2(0)  \cdot 2 \Theta_2 R_2^2 \pi    
\end{eqnarray}
\noindent We also have, 
\begin{eqnarray}
{\cal I}_{\theta_2 r_2 \theta_2} &=& 0 \\ 
F_{\theta_2\theta_3}&=& 0 \\
D_{\frac{\partial}{\partial \theta_3}} F_{r_{2}\theta_{2}} &=& 0 \\ 
D_{\frac{\partial}{\partial \theta_3}} F_{r_{2}\theta_{3}}\Big{|}_{r_3(0) = \pi/4 }  &=& 0 \\
F_{r_2\theta_2}  \cdot {\cal I}_{r_2\theta_2} &=& F_{\theta_2 r_2 } \cdot {\cal I}_{\theta_2 r_2}\\ 
F_{r_2\theta_3}\Big{|}_{\theta_3(0) = \pi} \cdot {\cal I}_{r_2\theta_3} &=&  F_{\theta_3 r_2}\Big{|}_{\theta_3(0) = \pi} \cdot {\cal I}_{\theta_3 r_2} \\
-2D_{\frac{\partial}{\partial r_2}} F_{r_{2}\theta_{3}}\Big{|}_{\theta_3(0) = \pi} \, \cdot {\cal I}_{r_2 r_2\theta_3}    &=&   
D_{\frac{\partial}{\partial r_2}} F_{\theta_{3} r_2}\Big{|}_{\theta_3(0) = \pi} \cdot {\cal I}_{r_2 \theta_3 r_2}   \\
2 D_{\frac{\partial}{\partial r_2}} F_{\theta_{2} r_2} \Big{|}_{\theta_2(0) = 0 }  \cdot  {\cal I}_{r_2   \theta_2 r_2}   &=& D_{\frac{\partial}{\partial r_2}} F_{r_{2}\theta_{2}} \Big{|}_{\theta_2(0) = 0 } \cdot  {\cal I}_{r_2 r_2  \theta_2} \, \, \, . 
\end{eqnarray}
The strategy now is to choose parameters so that the $F_{r_2\theta_2}$ terms kill the terms along the diagonal in the expressions
$D_{\frac{\partial}{\partial r_2}} F_{r_{2}\theta_{2}}$ and $D_{\frac{\partial}{\partial r_2}} F_{\theta_{2} r_2} \, .$  Then, with those parameters chosen, we choose the rest of the parameters so that
the remaining terms combine to yield $-i \theta I_{yy} \, , $ where $\theta$ is a free parameter.  Remembering to include the coefficients in the expansion,
the first objective leads to the following equation, 
\be\label{ob1}
r_2(0) = -\frac{1}{2}\mathrm{arctanh}( 2 R_2) \, . 
\ee

Setting $\Theta_2 = \Theta_3 = \theta \, , $ and substituting the previous equation defining $r_2(0)$ and $R_2 \, $,  the second objective yields
\be\label{nonzero}
R_2 = -\frac{\sin 2 n \pi }{4 \pi (n^2 - 4)} \, . 
\ee If $n$ is chose to as a non-integer so that (\ref{nonzero}), the loop $\gamma^*$ determines the holonomy (up to third order)
\be
\Gamma(\gamma^*) = \exp( -i \theta I_{yy}) \, 
\ee where $\theta$ is a free parameter.

Loops generating single qubit $SU(2)$ holonomies can be characterized by {\em abelianizing} the dynamics \cite{pach2}.
Thus with this single two-qubit holonomy, we can construct any $SU(4)$ transformation 
on the computational basis.  To see this, recall that any $SU(4)$ transformation may be written with the Cartan decomposition 
\be \label{decomp}
g = K_1 \exp (-i\theta_1 I_{xx}  - i \theta_2 I_{yy} - i \theta_3 I_{zz}  ) K_2 \, 
\ee where $K_j \in SU(2) \otimes SU(2) \, . $

We may obtain the transformations $\exp (-i\theta_1 I_{xx})$ and  $\exp (-i\theta_1 I_{zz})$  by noting that
\begin{eqnarray}
&&K_z^{-1}(\frac{\pi}{2})\exp(-i\theta I_{yy} )K_z(\frac{\pi}{2})  = \exp( -i\theta I_{xx} ) \\
&&K_x(\frac{\pi}{2})\exp(-i\theta I_{yy})K_x^{-1}(\frac{\pi}{2})  = \exp( -i\theta I_{zz} )  \, 
\eea
where
\bea
&&K_z(\theta) = \exp( - i \theta I_{z1}) \exp( - i \theta I_{z2} ) \\ 
&&K_x(\theta) = \exp( - i \theta I_{x1}) \exp( - i \theta I_{x1} ) \, . 
\eea

Thus any $g \in SU(4)$ can be approximated up to third order by, 
\be
g = K_1K_z^{-1}(\frac{\pi}{2}) \Gamma(\gamma^*)K_z(\frac{\pi}{2}) \Gamma(\gamma^*) K_x(\frac{\pi}{2})\Gamma(\gamma^*)K_x^{-1}(\frac{\pi}{2})K_2 
\ee
One can also use the {\em abelianization} procedure of the previous section to construct a 
holonomy of the form \cite{pach1,pach2}
\be 
\widetilde{U} = \frac{1}{\sqrt{2}}
\left(
\begin{matrix}
\sqrt{2} &  0& 0 &  0 \\ 0 & 1 & -i & 0 \\
0 & -i&1& 0 \\ 0 & 0 & 0 & \sqrt{2}
\end{matrix}
\right) \quad .
\ee 
One can then show that the Lie algebra element  $\tilde{\xi}$ such that $\exp(\tilde{\xi}) = \widetilde{U}$ 
along with the local algebra generates $\mathfrak{su}(4) \, $ under repeated bracketing.  Thus $\widetilde{U}$ 
is a so called {\em universal logic gate}.  However, this procedure does not give a prescription for 
building an arbitrary $SU(4)$ transformation.

\subsection{Conditional Berry Phases}
An interesting hybrid scheme to quantum computing involving {\em dynamical} $SU(2)$ rotations and 
conditional Berry phases has been realized as a universal set of gates for several physical systems
proposed for quantum computing.  To date, there have been HQC implementations using 
this control paradigm with NMR \cite{jones}, trapped ions \cite{duan}, neutral atoms \cite{recati}, 
semiconductor nanostructures \cite{solinas}, and Josephson junction networks  \cite{faoro,choi}.  
We refer the reader to the literature for a description of the physical systems 
underlying these proposed quantum computing schemes. 

Here we are interested in the control strategy of the experimentalist with the gates available in 
systems of this type.  Namely, how does one build an arbitrary unitary transformation on two coupled qubits 
given only single qubit rotations  and the conditional phase shift ?  It is perhaps surprising that indeed 
this is possible and one can entangle qubits with only the conditional phase shift as the non-local operation.  
For this model we do not concentrate on the generation of the fundamental logic gates since the 
$SU(2)$ transformations are typically not holonomic and the Abelian Berry phase contributing to the
conditional phase gate can be computed with Stokes theorem.

Since the conditional phase gate is not an element of $SU(4) \, ,$ we employ a Cartan decomposition of $U(4) \, . $
To this end, recall the notation of the previous section and note that  the real span of the sets
\bea \label{u(4)}
&&\mathfrak{k}  =  i \{ I_{x1}, \, I_{y1}, \, I_{z1}, \, I_{x2}, \, I_{y2}, I_{z2} \}   \\ 
&& \mathfrak{p} =  i \{{\bf 1}_4\, , I_{xx}, \, I_{yy}, \, I_{zz}, \, I_{xy}, \, I_{xz}, \, I_{yx}, \, I_{yz}, \, I_{zx}, I_{zy}    \}
\eea
form a basis of $\mathfrak{u}(4)$ in the tensor product representation.  Moreover, one can check the
commutation relations (\ref{cartan_comm}) to confirm that the set forms a Cartan decomposition of 
$\mathfrak{u}(4)$ where $\mathfrak{u}(4) = \mathfrak{k} \oplus \mathfrak{p} \, . $
Since the maximal Abelian subalgebra $\mathfrak{a} $ contained in $\mathfrak{p} $ is just 
\begin{equation}
\mathfrak{a} = i \{{\bf 1}_4\, , I_{xx}, \, I_{yy}, \, I_{zz}  \} \, , 
\end{equation} 
we obtain the decomposition for any $G \in U(4)$
\begin{equation}
G = K_1 \exp( -i\phi_0 {\bf 1}_4 -i\phi_1  I_{xx} \,-i\phi_2  I_{yy} \,-i\phi_3  I_{zz} )K_2
\end{equation}
where $\phi_i \in \mathbb{R} $ and $K_j \in K = SU(2) \otimes SU(2)  \, .$

\subsubsection{Control Algorithms}
Proceeding along the lines of \cite{dalles3,khan1,khan2}, we develop control algorithms 
with single qubit rotations and the conditional phase shift.   
The action of the conditional  phase  shift of the the computational basis is as follows,
\begin{eqnarray}
&&U_\phi | 00 \rangle = | 00 \rangle  , \quad  U_\phi \,  | 01 \rangle = | 01 \rangle  \\
&&U_\phi | 10 \rangle = | 10 \rangle  ,  \quad U_\phi \,  | 11 \rangle = e^{-i\phi} \, | 11 \rangle   \, .
\end{eqnarray}
Under the isomorphism $ C^2 \otimes C^2 \cong C^4 $, 
the conditional phase shift can be written  as
\begin{equation}
U_{\phi} = 
\left(
\begin{array}{cccc}
1&0&0&0\\
0&1&0&0\\
0&0&1&0 \\
0&0&0&e^{-i\phi}
\end{array}
\right) \, . 
\end{equation}
In terms of the basis (\ref{u(4)}) we can write
\begin{equation}
U_\phi = \exp(-i \frac{\phi}{2} (2{\bf 1}_4 + I_{zz} - (I_{z1} + I_{z2}))) \, .
\end{equation}
Let  \begin{equation}
\widetilde{K}_z = \exp( -i \frac{\phi}{4} ( I_{z1} + I_{z2}) )\, , 
\end{equation}
since
\begin{equation}
[ 2 {\bf 1}_4 , I_{zi}] =  [ I_{zz} , I_{zi} ] = 0 
\end{equation}
we have 
\begin{eqnarray}
\widetilde{K}_z U_{\phi/2} &=& \exp( -i \frac{\phi}{4} ( I_{z1} + I_{z2}))  \nonumber 
\cdot \exp(-i\frac{\phi}{4} (2 {\bf 1}_4 + I_{zz}  - (I_{z1} + I_{z2})))  \nonumber  \\
&=& \exp(-i \frac{\phi}{4} (2{\bf 1}_4 + I_{zz} )) \, . 
\end{eqnarray}
Let  \begin{equation}
K_{j} (\theta) = \exp(-i \theta  I_{j 1 } ) \exp(- i \theta I_{j 2} )  \in K
\end{equation}
for $j  \in \{ x, \, y \}  \, $ and $\theta \in \mathbb{R}$ is a real parameter.   We have the commutation relations,
\be
{\big [} -i( I_{z1} + I_{z2}) \, , \,  -i(2 {\bf 1}_4 + I_{zz}) {\big ]} =   {\big [} -i( I_{z1} + I_{z2})\, , \, -i I_{zz} {\big ]} \ee
and 
\be
{\big [} -iI_{zz} \,  , \, [ -i(I_{j1} + I_{j2})\, , \, -i I_{zz} ] {\big]} = -i(I_{j1} + I_{j2}) \, . 
\ee
Thus by the Campbell-Baker-Hausdorff formula, we obtain, 
\bea\label{decouple}
K_{j} (\theta) \widetilde{K}_z U_{\phi / 2} K_{j}^{-1}(\theta)   & = &  K_{ j}  (\theta) \exp(-i\frac{\phi}{4}(2 {\bf 1}_4 + I_{zz} )) K_{j}^{-1}  (\theta)  \\
& = & \exp(-i\frac{\phi}{4} (2 {\bf 1}_4 + I_{zz} \cos(\theta)   + {\big[} (- i I_{j1} - i I_{j2}) , -i I_{zz} {\big]} \sin(\theta) ))  \, . 
\eea
We employ a $\pi$-rotation to achieve the  necessary decoupling.  Using the preceding  expression we get,
\bea
K_{j } (\pi) \widetilde{K}_z U_{\phi/2} K_{j}^{-1}(\pi)  &=&  K_{j} (\pi) \exp(-i\frac{\phi}{4} (2{\bf 1}_4 + I_{zz} )) K_{j}^{-1}  (\pi)  \\
&=& \exp(-i\frac{\phi}{4} (2{\bf 1}_4 -  I_{zz} ))  \, .
\eea
So we obtain 
\begin{equation}
\widetilde{K}_z U_{\phi /2}  \cdot K_{ j}  (\pi) \widetilde{K}_z U_{\phi / 2}  K_{ j}^{-1}(\pi)  = \exp(-i\phi {\bf 1}_4) \, . 
\end{equation}
Similarly, 
\begin{equation}
\widetilde{K}'_z U_{\phi} \cdot   K_j (\pi) \widetilde{K}'_z U_{-\phi}  K_j^{-1}(\pi)  = \exp(-i\phi I_{zz}) \,  
\end{equation}
where $\widetilde{K}'_z = \exp( -i \frac{\phi}{2} ( I_{z1} + I_{z2}) ) \, . $

Finally, by noting that for 
\bea
&&K_x(\theta) = \exp( - i \theta I_{x1})  \exp( - i \theta I_{x2} ) \\ 
&&K_y(\theta) = \exp( - i \theta I_{y1})  \exp( - i \theta I_{y1} ) 
\eea
we have 
\begin{eqnarray}
&K_y(\frac{\pi}{2})\exp(-i\phi I_{zz} )K_y^{-1}(\frac{\pi}{2})  = \exp( -i\phi I_{xx} )  \label{flip1}\quad \\
&K_x^{-1}(\frac{\pi}{2})\exp(-i\phi I_{zz})K_x(\frac{\pi}{2})  = \exp( -i\phi I_{yy} ) .  \label{flip2} \quad  
\end{eqnarray}
Using (\ref{flip1})and (\ref{flip2}), we can construct the desired decomposition for any $G \in U(4) \, ,$
\begin{eqnarray}
G &=& K_1\cdot \exp( -i\phi_0 {\bf 1}_4 -i\phi_1  I_{xx} -i\phi_2  I_{yy} -i\phi_3  I_{zz} )  \cdot K_2 \\
 &=&  K_1\cdot  \exp( -i \phi_0 {\bf 1}_4 ) \cdot  K_y(\frac{\pi}{2}) \exp(-i\phi_1 I_{zz} ) K_y^{-1} (\frac{\pi}{2})  \cdot 
  K_x^{-1}(\frac{\pi}{2})  \exp(-i\phi_2 I_{zz}) K_x (\frac{\pi}{2}) \\  \nonumber
  && \quad \quad  \cdot \exp(-i\phi_3 I_{zz} ) \cdot  K_2  \,  . 
\end{eqnarray}
This can now be written in terms of just elements of $SU(2) \otimes SU(2)$ and the conditional phase shift $U_\phi \, , $ 
\begin{eqnarray}
G &= K_1  \cdot \widetilde{K}_z U_{\phi_0 / 2} K_{j} (\pi) \widetilde{K}_z U_{\phi_0 / 2} K_{j}^{-1}(\pi) \cdot  K_y(\frac{\pi}{2})  \widetilde{K}'_z 
 U_{\phi_1} K_{j} (\pi) \widetilde{K}'_z U_{-\phi_1}  K_{j}^{-1}(\pi)  K_y^{-1} (\frac{\pi}{2}) \cdot  K_x^{-1} (\frac{\pi}{2}) 
 \nonumber \\ & \, \, \, \, \, \, \, \widetilde{K}'_z U_{\phi_2} K_{j} (\pi) \widetilde{K}'_z U_{-\phi_2}  K_{j}^{-1}(\pi)  K_x (\frac{\pi}{2})  
  \widetilde{K}'_z U_{\phi_3}  K_{j} (\pi) \widetilde{K}'_z U_{-\phi_3}  K_{j}^{-1}(\pi) \, \cdot \, K_2 \,. 
\end{eqnarray}
Given the freedom of choosing $j$ in $K_{j}$ this sequence can be simplified $\mathrm{somewhat .} \, $  
For example,  choose $ \, j = y \, $ in the product $ K_{j }^{-1}(\pi)  \, \cdot \, K_y(\frac{\pi}{2}) $ to obtain 
$K_{j }^{-1}(\pi)  \, \cdot \, K_y(\frac{\pi}{2})=K_{y }(-\pi)  \, \cdot \, K_y(\frac{\pi}{2})=K_y(-\frac{\pi}{2})  \, . $
Using this substitution and  two others, the decomposition simplifies to 
\begin{eqnarray}
G &=  K_1 \cdot \widetilde{K}_z U_{\phi_0 / 2} K_{j} (\pi) \widetilde{K}_z U_{\phi_0 / 2}  K_y(-\frac{\pi}{2})  \widetilde{K}'_z U_{\phi_1}  K_{j} (\pi) 
 \widetilde{K}'_z U_{-\phi_1}  K_{y}(-\frac{3\pi}{2})  K_x^{-1} (\frac{\pi}{2}) \widetilde{K}'_z U_{\phi_2}  K_{j} (\pi) \widetilde{K}'_z U_{-\phi_2} 
 \nonumber \\ &  K_x(-\frac{\pi}{2})  \, \cdot \, \widetilde{K}'_z U_{\phi_3}  K_{j} (\pi)   \widetilde{K}'_z U_{-\phi_3}  K_{j}^{-1}(\pi) \cdot K_2 \, . 
\end{eqnarray}
Finally, absorbing  $\widetilde{K}_z$ and $ K_{j}^{-1}(\pi) $ into $K_1$ and $K_2$ respectively, we get
\begin{eqnarray}
G &=  K_1 U_{\phi_0 / 2} K_{j} (\pi) \widetilde{K}_z U_{\phi_0 / 2}  K_y(\frac{\pi}{ 2})\widetilde{K}'_z U_{\phi_1}  K_{j} (\pi) \widetilde{K}'_z
  U_{-\phi_1}  K_{y}(-\frac{3\pi}{2})  K_x^{-1} (\frac{\pi}{2}) \widetilde{K}'_z U_{\phi_2}  K_{j} (\pi) \widetilde{K}'_z U_{-\phi_2} 
 \nonumber \\&K_x (- \frac{\pi}{2})  \widetilde{K}'_z U_{\phi_3}  K_{j} (\pi) \widetilde{K}'_z U_{-\phi_3}   K_2 \,. 
\end{eqnarray}
Some remarks are appropriate.   This sequence of unitary transformations is exact and a precise 
prescription for building any $U(4)$ logic gate with just local operations and the conditional phase shift.  
We make no claim that this decomposition is optimal with respect to number of elements nor time.  In the holonomic 
framework, time optimality is constrained by the adiabatic requirement.  In this case, one should 
then focus primarily on minimizing the number of loops necessary to build an arbitrary gate.

\section{Conclusion}
In this paper, we have considered holonomic quantum computation from a control theoretic point of view.  A general framework
for the control analysis is obtained by casting the relevant problems as control systems evolving on principal fiber bundles.  We have applied 
this framework to two well established models of the computing scheme.  To the best of our knowledge, all holonomic computing schemes
proposed thus far fall into one of the two models considered here.  From a control perspective, an interesting avenue for future work would be
extending these ideas to the control of molecular systems in the Born-Oppenheimer approximation (as mentioned in \cite{zan}).   Holonomies
can be realized in this regime \cite{bohm} and it is reasonable to expect that a similar analysis can be carried out for these systems.  However, 
a direct application of the methods proposed here will not suffice since the control parameters themselves are quantum degrees of freedom 
and therefore possess a non-trivial uncontrolled evolution of their own.  In other words,  the state equations analogous to those
considered here (\ref{CSPB}) will be  coupled quantum control problems.   

\section{Acknowledgments}
The author thanks Professors T.~J. Tarn and John Clark for their guidance and support.  The author
gratefully acknowledges the financial support of Washington University in St. Louis where the majority of this research was
conducted. The completion of this work was supported in part by JHU/APL part-time study funds.

\newpage
\appendix
\begin{center}
\bf{\Large{Appendix}}
\end{center}
\section{Two-Qubit Connection Components}
\bea
A_{r_{2}} &=&
\left(
\begin{matrix} 0 & 0 & 0 & -e^{-i\theta_2} \\
0 & 0 & 0 & 0 \\ 0 & 0 & 0 & 0 \\ e^{i\theta_2}& 0 & 0 & 0
\end{matrix}
\right)  \notag  \\ && \notag \\
A_{r_{3}} &=&
\left(
\begin{matrix} 0 & 0 & 0 & 0 \\ 0 & 0 & -e^{-i\theta_3} & 0 \\
0 & e^{i\theta_3} & 0 & 0 \\ 0 & 0 & 0 & 0 \\
\end{matrix}
\right) (2\cosh^2r_2 - 1) \notag 
\eea
\bea
A_{\theta{_2}}  \, &=& \,
\left(
\begin{matrix}
0 & 0 & 0 & e^{-i\theta_2} \\
0 & 0 & 0 & 0 \\0 & 0 & 0 & 0 \\ e^{i\theta_2}& 0 & 0 & 0
\end{matrix}
\right) \frac{i}{2} \sinh2r_2 \, + \,
\left(
\begin{matrix}
1 & 0 & 0 & 0 \\
0 & 2 & 0 & 0 \\ 0 & 0 & 2 & 0 \\ 0 & 0 & 0 & 3
\end{matrix}
\right)  \frac{i}{2} ( \cosh2r_2 - 1 ) \,  \notag  \\ && \notag  \\ 
A_{\theta{_3}} \, &=& \,
\left(
\begin{matrix}
0 & 0 & 0 & 0 \\ 0 & 0 & e^{-i\theta_{3}} & 0 \\
0 & e^{i\theta{3}} & 0 & 0 \\ 0 & 0 & 0 & 0
\end{matrix}
\right)
\frac{i}{2} \cosh 2r_2 \sin 2r_3 \, + \,
\left(
\begin{matrix}
0 & 0 & 0 & 0 \\ 0 & 1 & 0 & 0 \\ 0 & 0 & -1 & 0 \\ 0 & 0 & 0 & 0
\end{matrix}
\right)
i \sin^2 r_3   \quad .  \notag 
\eea
\section{Interaction Holonomy Algebra}
\bea 
F_{r_{2}r_{3}} \,&=&  \,
\left(
\begin{matrix} 0 & 0 & 0 & 0 \\
0 & 0 & -e^{-i\theta_3} & 0 \\ 0 & e^{i\theta_3} & 0 & 0 \\ 0& 0 & 0 & 0
\end{matrix}
\right)  2 \, \sinh 2r_2   \notag \\ && \notag \\
F_{r_{2}\theta_{2}} \, &=&  \,
\left(
\begin{matrix} 0 & 0 & 0 & 0 \\ 0 & 1 & 0 &  0 \\
0 &  0 & 1 & 0 \\ 0 & 0 & 0 & 2 \\
\end{matrix}
\right)
2 i \,  \sinh 2r_2 \,  \notag  \\
F_{r_{2}\theta_{3}}   \, &=&  \,
\left(
\begin{matrix}
0 & 0 & 0 & 0 \\
0 & 0 & e^{-i \theta_{3}} & 0 \\ 0 & e^{i \theta_3}& 0 & 0 \\ 0& 0 & 0 & 0 \\
\end{matrix}
\right)  \,
i \sin 2r_3 \sinh 2r_2  \notag \\ && \notag 
\eea
\be
F_{r_{3}\theta_{3}} \, =  \,
\left(
\begin{matrix} 0 & 0 & 0 & 0 \\ 0 & -1 & 0 &  0 \\
0 &  0 & 1 & 0 \\ 0 & 0 & 0 & 0 \\
\end{matrix}
\right)
 i \,  \sin 2 r_3 \sinh^2 2r_2 \, ,  \notag 
\ee
\begin{eqnarray*}
&& D_{\frac{\partial}{\partial \theta_2}} F_{r_{2}\theta_{2}} \, =   \,
\left(
\begin{matrix}
0 & 0 & 0 &  - e^{-i\theta_2} \\ 0 & 0 & 0 & 0 \\
0 & 0 & 0 & 0 \\ e^{i \theta_2} & 0 & 0 & 0
\end{matrix}
\right)
 2 \sinh^2 2 r_2    \\&& \\ 
&&D_{\frac{\partial}{\partial r_2}} F_{r_{2}\theta_{2}} \, =  \,
\left(
\begin{matrix}
0 & 0 & 0 &  -e^{-i\theta_2} \\ 0 & 0 & 0 & 0 \\
0 & 0 & 0 & 0 \\ -e^{i \theta_2} & 0 & 0 & 0
\end{matrix}
\right)
4i\sinh 2r_2
\, +  \left(
\begin{matrix} 0 & 0 & 0 & 0 \\ 0 & 1 & 0 &  0 \\
0 &  0 & 1 & 0 \\ 0 & 0 & 0 & 2 \\
\end{matrix}
\right)
4i \cosh 2r_2 \, ,  \\ && \\
&&D_{\frac{\partial}{\partial \theta_2}} D_{\frac{\partial}{\partial \theta_2}}
F_{r_{2}\theta_{2}} \, =  \,
\left(
\begin{matrix}
1 & 0 & 0 & 0 \\ 0 & 0 & 0 & 0 \\ 0 & 0 & 0 & 0 \\ 0 & 0 & 0 & -1
\end{matrix}
\right)
2 i \sinh^3 r_2 \, + \left(
\begin{matrix}
0 & 0 & 0 & e^{-i\theta_2} \\ 0 & 0 & 0 & 0 \\
0 & 0 & 0 & 0 \\ e^{i \theta_2} & 0 & 0 & 0
\end{matrix}
\right)  \,
2 i \sinh^2 2r_2 \cosh 2r_2   
\end{eqnarray*}
\section{Higher Order Brackets}
\begin{eqnarray*}
[\,  I_{x1} \, , \, F_{r_2 r_3} \, ] &=&
\left( 
\begin{matrix}
0&e^{i\theta_3}&0&0 \\ 
e^{-i\theta_3}&0&0&0 \\
0&0&0&-e^{i\theta_3} \\
0&0&-e^{-i\theta_3}&0 \\
\end{matrix}
\right)i \sinh 2 r_2  \\ 
\left[  I_{x1} \, , \, F_{r_2 \theta_3} \right] &=&
\left( 
\begin{matrix}
0&-e^{i\theta_3}&0&0 \\ 
e^{-i\theta_3}&0&0&0 \\
0&0&0&e^{i\theta_3} \\
0&0&-e^{-i\theta_3}&0 \\
\end{matrix}
\right)\frac{1}{2}\sin 2 r_2 \sinh 2 r_2 \\&&\\
\left[ I_{x2} \, , \, F_{r_2 r_3}  \right] &=&
\left( 
\begin{matrix}
0&0&-e^{-i\theta_3}&0 \\ 
0&0&0&e^{-i\theta_3} \\
-e^{i\theta_3}&0&0&0 \\
0&e^{i\theta_3} &0&\\
\end{matrix}
\right)i \sinh 2 r_2 \\&&\\
\left[  I_{x2} \, , \, F_{r_2 \theta_3} \right] &=&
\left( 
\begin{matrix}
0&0&-e^{-i\theta_3}&0 \\ 
0&0&0&e^{-i\theta_3} \\
e^{i\theta_3}&0&0&0 \\
0&-e^{i\theta_3} &0&\\
\end{matrix}
\right)\frac{1}{2}\sin 2 r_2 \sinh 2 r_2 \\&&\\
\left[  [\,  I_{x1} \, , \, F_{r_2 r_3} \, ]    \, , \, I_{x2}\right] &=&
\left(
\begin{matrix}
-1&0&0&\\0&1&0&0\\0&0&1&0\\0&0&0&-1
\end{matrix}
\right)i \sin \theta_3 \sinh 2 r_2
\end{eqnarray*}

\end{document}